\documentclass[%
 reprint,
nofootinbib,
 amsmath,amssymb,
 aps,
]{revtex4-2}

\usepackage{physics}
\usepackage[pdftex]{graphicx}
\usepackage{dcolumn}
\usepackage{bm,float}
\usepackage{mathtools}
\usepackage{mathrsfs}
\usepackage{hyperref}
\hypersetup{
setpagesize=false,
bookmarksnumbered=true,
bookmarksopen=true,%
colorlinks=true,%
linkcolor=red,
citecolor=blue,
}

\usepackage{ulem}
\usepackage{xcolor}
\DeclareRobustCommand{\erase}{\bgroup\markoverwith{\textcolor{red}{\rule[.5ex]{2pt}{0.8pt}}}\ULon}

\usepackage{cancel}

\date{\today}

\begin{abstract}
We investigate the entropy dynamics of de Sitter spacetime during the inflationary phase. The cosmological horizon in de Sitter spacetime, which limits the causally accessible region for an observer, exhibits thermal properties similar to a black hole event horizon. According to holographic principles, the entropy within a causally connected region is bounded by its surface area. However, this entropy bound is violated during the eternal phase of inflation.
To address these violations from a quantum  information perspective, we adopt a stochastic approach to cosmic inflation. Specifically, we analyze the Shannon entropy of the inflaton field’s probability distribution, which mirrors the behavior of the entanglement entropy of a Hubble-sized region in stochastic inflation. Using the volume-weighted probability distribution for the inflaton field, we demonstrate a significant entropy behavior in de Sitter spacetime.
\end{abstract}

\begin{document}
\title{Stochastic inflation and entropy bound in de Sitter spacetime}

\author{Hiromasa Tajima}
 \email{tajima.hiromasa.m6@s.mail.nagoya-u.ac.jp}
\author{Yasusada Nambu}%
\email{nambu.yasusada.e5@f.mail.nagoya-u.ac.jp}
\affiliation{%
 Department of Physics, Graduate School of Science, Nagoya University, Nagoya 464-8602, Japan
}%

\maketitle


\section{\label{sec:Intro}Introduction}

Understanding quantum gravity requires a thorough assessment of the theory's effectiveness.
Entropy consistently plays a crucial role in our research approach.
According to the holographic principle, the entropy within a region is bounded by the area of its boundary.
The entropy bound was originally conjectured by Bekenstein~\cite{Bekenstein:1980jp} and later refined by Bousso~\cite{Bousso:1999xy}.
The violation of the entropy bound is related to the information problem.
For example, consider the case of black holes that evaporate at late times.
The violation of the entropy bound in black hole spacetimes is called the information loss paradox~\cite{Hawking:1975vcx}.
Conversely, the violation of the entropy bound in the de Sitter spacetime indicates that a long de Sitter phase during inflation is in swampland~\cite{Arkani-Hamed:2007ryv}.
The violation of the entropy bound has been discussed from the perspective of quantum information~\cite{Kames-King:2021etp, Geng:2021wcq}

%
An observer in a de Sitter spacetime cannot observe an entire spacetime.
This is analogous to the black hole case, where the event horizon hides a finite region of spacetime from external observers.
The boundary that restricts an observer's accessible region is termed the ``cosmological horizon" in de Sitter spacetime.
We can assign the temperature and entropy of the cosmological horizon to be similar to the event horizon of black holes~\cite{Gibbons:1977mu}.
In \cite{Arkani-Hamed:2007ryv}, the violation of the entropy bound in the de Sitter spacetime was treated with an analogy to the black hole case. 

Based on the analogy with black holes, we posit that the quantum gravity effects are similarly manifested in the de Sitter spacetime.
Recent advances in resolving the black hole information paradox~\cite{Page:1993wv, Penington:2019npb, Almheiri:2019hni, Penington:2019kki} suggest that spacetime should be globally considered when accounting for the quantum gravity effects.
The appearance of an island region indicates the global effects of quantum gravity~\cite{Penington:2019npb, Almheiri:2019hni, Penington:2019kki}.
Existing approaches have used quantum information concepts to study the effects of quantum gravity, such as dS/CFT correspondence~\cite{Balasubramanian:2020xqf, Chen:2020tes, Geng:2021wcq, Teresi:2021qff, Hikida:2022ltr, Kawamoto:2023nki}.
However, there is currently no guiding principle equivalent to the Page curve~\cite{Page:1993wv}  adopted in black hole cases to study the quantum gravity effects in the de Sitter spacetime from a quantum information perspective.

First, this study reformulates the e-folding bound issue \cite{Arkani-Hamed:2007ryv} in eternal inflation, linking it to the entropy bound and its violation. We adopt a stochastic approach to de Sitter spacetime during inflation to address the information problem~\cite{Starobinsky:1986fx, Rey:1986zk, Vilenkin_1983}.
We focus on the superhorizon modes of the inflaton field while coarse-graining the subhorizon modes within the horizon-sized region (h-region). Due to the classicalization of quantum fluctuations resulting from the de Sitter expansion of the universe, the quantum system of superhorizon modes can be effectively described as a classical stochastic system, leading to the classicalization of quantum correlations.
As a result, we find that the entanglement entropy (EE) of a specified spatial region can be expressed as the Shannon entropy of the inflaton field’s probability distribution through the stochastic approach. This allows us to reinterpret the entropy bound proposed by Arkani-Hamed \textit{et al.} \cite{Arkani-Hamed:2007ryv} as an inequality governing the amount of information. In particular, we demonstrate that the inequality in \cite{Arkani-Hamed:2007ryv} is equivalent to the information inequality derived from the Bousso bound \cite{Bousso:2006ge}. The violation of this inequality, which Arkani-Hamed \textit{et al.} claimed in the context of eternal inflation, also arises in our framework, leading us to conclude that this violation represents the information problem in de Sitter spacetime.

Next, to address the information problem we identify, we derive the guiding behavior of entropy, which is analogous to the Page curve in the context of the black hole information loss paradox. In the black hole case, the global black hole state is treated as a random pure state to derive the Page curve—an operation equivalent to applying meaningful weightings to the system’s probability distribution.
Moreover, the effect of global expansion is crucial when considering the global structure of eternal inflation. For these reasons, it is necessary to apply volume weighting to obtain the typical state in de Sitter spacetime.

The remainder of this paper is organized as follows.
In Section~\ref{sec:ent_bound}, we review the proposal of the e-folding bound from Arkani-Hamed \textit{et al.}~\cite{Arkani-Hamed:2007ryv}
and the entropy bound during the inflationary universe.
In Section~\ref{sec:SI_Ent}, we introduce the stochastic approach to de Sitter spacetime. We then discuss the conditions for classicalization and examine the entropy bound defined by \cite{Arkani-Hamed:2007ryv} within our framework.
In Section~\ref{sec:dw_pte_inf}, we investigate the information problem in de Sitter (dS) spacetime within our setup. Additionally, we explore volume-weighting as a physically meaningful state-counting method, analyze its associated entropy bound, and assess the impact of modifying the weighting.
Finally, Section~\ref{sec:sum_conc} summarizes our findings.

\section{\label{sec:ent_bound}ENTROPY BOUND FOR INFLATIONARY UNIVERSE}
\subsection{Operational meaning of Gibbons-Hawking entropy}

We review the entropy bound in the inflationary universe and its operational meaning discussed in~\cite{Arkani-Hamed:2007ryv}.
First, we will introduce the setup of single-field slow-roll inflation with a stochastic approach and the entropy bound in the de Sitter spacetime during inflation introduced in \cite{Arkani-Hamed:2007ryv}.

We consider inflation in the Friedman-Lema\^{i}tre-Robertson-Walker (FLRW) universe as follows:
\begin{align}
  \dd{s}^2=-\dd{t}^2+a^2\qty(t)\dd{\boldsymbol{x}}^2,\quad a\propto\exp(N(t)) ,
\end{align}
where the e-folding number is defined by
\begin{equation}
N(t)=\int ^t dt\, H(t),\quad H(t)=\frac{\dot a(t)}{a(t)},
\end{equation}
and the dot on the scale factor denotes the derivative with respect to time $t$.
In the slow-roll inflation, the inflaton field $\varphi$ in the Hubble-size region (h-region) obeys the following equation:
\begin{align}
  \frac{d\varphi}{dt}=-\frac{V'(\varphi)}{3H(\varphi)},
  \label{inflaton_eom_classical}
\end{align}
where the dash denotes the derivative with respect to $\varphi$ and the Hubble parameter $H(\varphi)$ is related to the inflaton potential $V\qty(\varphi)$ by the  Friedmann equation
\begin{align}
  H^2\qty(\varphi)\simeq \frac{\kappa}{3}V\qty(\varphi),\quad \kappa:=8\pi G = \frac{8\pi}{m_p^2}.
\end{align}

Associated with the de Sitter horizon, the Gibbons-Hawking (GH) entropy \cite{Gibbons:1977mu} is defined by
\begin{align}
    S_{\mathrm{GH}}
    =\frac{8\pi^2}{\kappa\hbar H^2}
    = \frac{24\pi^2}{\kappa^2\hbar V},
    \label{eq:GH}
\end{align}
and corresponds to the Bekenstein-Hawking entropy for a black hole. In other words, $S_{\mathrm{GH}}$ bounds the entropy of the h-region in de Sitter spacetime.
In our setup, an observer observes the spacetime region bounded by the past light cone after the end of inflation, as shown in Fig.~\ref{fig:dS_observer}.
The observer can observe many h-regions on the reheating surface.
They can detect the energy density $\rho$ and its fluctuation $\delta\rho$ or the gauge invariant curvature fluctuation $\mathcal{R}$. The density fluctuation originates from the quantum fluctuation of the inflaton field that causes curvature fluctuation $\mathcal{R}$ and induces the density fluctuation $\delta\rho/\rho\sim \mathcal{R}$ on the flat slice at the horizon re-entry after inflation.

Here, we consider the fluctuation at the scale of an h-region, which obeys
\begin{align}
    \expval{\mathcal{R}^2}
    = \hbar
    \left(
    \frac{H^2}{2\pi\dot{\varphi}}
    \right)^2=
    \frac{\hbar\kappa^3}{12\pi^2}
    \frac{V^3(\varphi)}{(V'(\varphi))^2},
\end{align}
where this quantity is evaluated when the wavelength of the fluctuation that leaves the Hubble horizon (horizon exit).

From the discussion of \cite{Arkani-Hamed:2007ryv},
we take the time derivative of the Gibbons-Hawking (GH) entropy $S_{\mathrm{GH}}$ with respect to the e-folding number $N$ is 
\begin{align}
    \frac{\dd S_{\mathrm{GH}}}{\dd N} = \frac{8\pi^2\dot{\varphi}^2}{\hbar H^4} = \frac{2}{\langle\mathcal{R}^2\rangle}.
\end{align}
Integrating this equation from the initial time $t_0$ to the end time of inflation $t_\text{end}$, we obtain
\begin{align}
    N\qty(t_\text{end})-N\qty(t_0)
    =   \frac{1}{2}\int^{S_{\mathrm{GH}}(t_\text{end})}_{S_{\mathrm{GH}}(t_0)}\dd{S_{\mathrm{GH}}}\langle\mathcal{R}^2\rangle.
\end{align}
In this study, we define $t_0$ as the moment at which the physical wavelength of perturbation crosses the Hubble horizon (horizon exit time. See Fig. \ref{fig:Super_horizon_mode}).
Then, the right-hand side of this equation can be rewritten as 
\begin{equation}
    N(t_\text{end})-N(t_0)
    = \frac{\langle\mathcal{R}^2\rangle_{t_*}}{2}\left[
    S_{\mathrm{GH}}\qty(t_\text{end})-S_{\mathrm{GH}}\qty(t_0)\right],
\end{equation}
where $t_*$ is a fixed value between $t_0$ and $t_\text{end}$ (the mean value theorem for definite integrals).
For the slow-roll inflation, as $\left<\mathcal{R}^2\right>$ decreases monotonically with time under the classical evolution of the inflaton field, $\left<\mathcal{R}^2\right>_{t_*}<\left<\mathcal{R}^2\right>_{t_0}$. Thus, we have the following inequality for the total e-folding number of inflation $\mathcal{N}_{\mathrm{e}}=N\qty(t_{\mathrm{end}})-N\qty(t_0)$:
\begin{align}
\mathcal{N}_e&\le\frac{\left<\mathcal{R}^2\right>_{t_0}}{2}
    \bigl[
    S_{\mathrm{GH}}\qty(t_\text{end})-S_{\mathrm{GH}}\qty(t_0)\bigr]\notag \\
    &\le\frac{\left<\mathcal{R}^2\right>_{t_0}}{2}
    S_{\mathrm{GH}}(t_\text{end}).
    \label{eq:entropy_bound}
\end{align}
As the number of inflaton modes crossed the cosmological horizon during inflation is roughly estimated as the volume of inflating region $\sim e^{3N(t)-3N(0)}$,
 this inequality provides the precise bound on the number of e-folds of the inflaton modes in terms of the GH entropy. We consider $t_0$ as the horizon exit time of the fluctuation.
In \cite{Arkani-Hamed:2007ryv}, the end time of inflation is defined by the time that the slow-roll parameter  becomes order unity:
\begin{align}
    \left|
    \frac{1}{\kappa}\frac{V''(\varphi)}{V(\varphi)}\right|_{t_\text{end}}\sim 1.
    \label{eq:inflaton_end_time}
\end{align}
The total e-folding number expresses the period of the de Sitter epoch during inflation. 
Therefore, they concluded that inflation cannot last too long because the entropy bound is violated. 
This violation will persist for the eternal phase of inflation, as the stochastic jump due to quantum fluctuations overwhelms the classical drift due to the potential force.
This corresponds to the concept of the eternal inflation phase from a local perspective. However, the eternal inflation phase is the phase for the global spacetime structure.
This is why we should consider the concept of an eternal inflation phase from a global perspective.

\subsection{Holographic Entropy Bound during Inflation with Stochastic Approach}
\begin{figure}[b]
\centering
\includegraphics[width=0.6\linewidth]{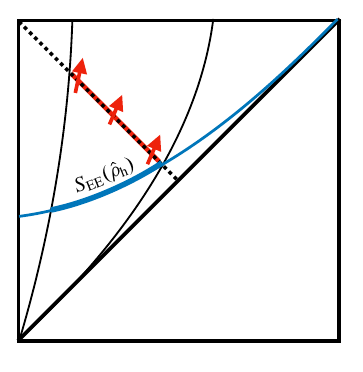}
\caption{
Holographic entropy bound in the de Sitter universe (Bousso’s entropy bound): The de Sitter horizon (dotted line) is a null surface that bounds the entropy flux (red arrows) emerging from it. The inequality \eqref{eq:Bousso_Bound} constrains the entanglement entropy (EE) of the h-region (blue line).}
\label{fig:dS_observer}
\end{figure}

From the holographic principle, the area of the boundary for the finite region in spacetime provides the maximum entropy in the region. This assumption is formulated by the covariant version so-called the Bousso bound~\cite{Bousso:2006ge}.
Regarding the de Sitter horizon as a light sheet, the Bousso 
bound in our setup is defined as
\begin{equation}
    S_{\mathrm{EE}}[\hat{\rho}_{\mathrm{h}}]\le S_{\mathrm{GH}},
    \label{eq:Bousso_Bound}
\end{equation}
where the $\hat{\rho}_{\mathrm{h}}$ is the density matrix of an h-region as shown in Fig.~\ref{fig:dS_observer}. This equation means that the maximum information is the GH entropy. Therefore, this inequality does not originate from inflation theory but from the discussion of information.
In the results of this paper, we demonstrate that Eq. \eqref{eq:entropy_bound}, and Eq. \eqref{eq:Bousso_Bound} exhibit remarkable equivalence within the framework of the stochastic formalism, despite the preceding discussion highlighting their fundamentally different origins.

Within the framework of stochastic inflation, the superhorizon mode of the inflaton field is defined as the mode that traverses the cosmological horizon and remains delocalized from any h-region. It is postulated that the superhorizon mode exhibits maximal quantum entanglement with the subhorizon mode present in an h-region. This assumption is made because the correlation among h-regions diminishes exponentially during expansion. Consequently, for the superhorizon mode's state $\hat{\rho}_{\mathrm{long}}$, the entanglement entropy (EE) within an h-region is characterized as
\begin{equation}
    S_{\mathrm{EE}}[\hat{\rho}_{\mathrm{long}}]
    =S_{\mathrm{EE}}[\hat{\rho}_{\mathrm{h}}].
\end{equation}
\begin{figure}[b]
\centering
\includegraphics[width=0.8\linewidth]{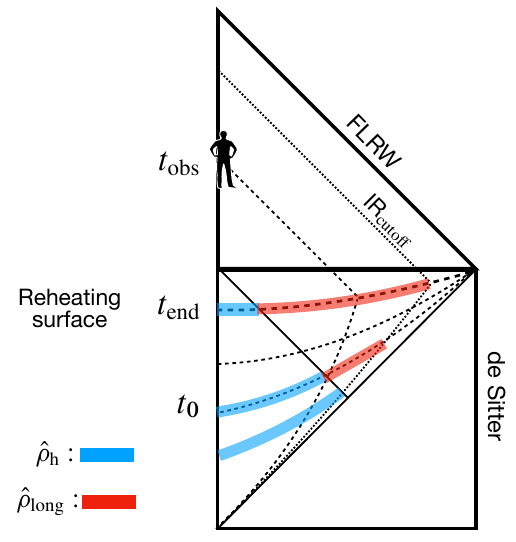}
\caption{
Information about the inflationary universe becomes accessible to an observer in the FLRW universe after inflation. Before inflation, spacetime is a pure de Sitter (dS) spacetime. During inflation, it transitions to a quasi-dS phase. After sufficient inflation, reheating occurs, ending inflation.
Superhorizon inflaton modes that crossed the horizon during inflation at $t_0$ later re-enter the particle horizon and become observable.
 Entanglement between superhorizon (red lines) and subhorizon modes  (blue lines) arises from their effective interactions. To compute EE, we set the infrared (IR) cutoff of the state for the physical mode  at the scale corresponding to the start of inflation, as the pre-inflationary phase is beyond our scope. }
\label{fig:Super_horizon_mode}
\end{figure}
In Eqs.~\eqref{eq:Ssh} and \eqref{eq:SEEnu}, we will confirm that the entanglement entropy $S_\text{EE}$ and the Shannon entropy $S_\text{Sh}$ for the inflaton in h-region are given as the logarithm of the symplectic eigenvalue $\nu\gg 1$ of the inflaton state.
The Shannon entropy quantifies the amount of classical information present in classical stochastic systems, and in our system, quantum information becomes equivalent to classical information.
Therefore, the Bousso bound~\eqref{eq:Bousso_Bound} becomes
\begin{align}
    S_{\mathrm{EE}}[\hat{\rho}_{\mathrm{long}}] 
    \approx S_{\mathrm{Sh}}[P(\varphi)]\leq S_{\mathrm{GH}},
    \label{entropy_bound_shannon}
\end{align}
where $P(\varphi)$ represents the probability distribution of the inflaton field in the h-region.
Because the entanglement entropy diverges by counting the entanglement generated before inflation, we should introduce the IR cutoff  by limiting the mode as in Fig.~\ref{fig:Super_horizon_mode}. 
The amount of entanglement during the finite period $t_0<t<t_*$ with this cutoff follows the modified version of Eq. \eqref{entropy_bound_shannon}: 
\begin{align}
S_{\mathrm{Sh}}&(t_*;t_0)
:=
S_{\mathrm{Sh}}(t_*)-S_{\mathrm{Sh}}(t_0)\leq S_{\mathrm{GH}}(t_*).
\label{entropy_bound_precise}
\end{align}
This is the inequality for information during inflation.

\section{Stochastic Approach to de Sitter Spacetime and its entropy}
\label{sec:SI_Ent}
Adopting the stochastic approach, the inflaton $\varphi$ obeys the following Langevin equation~\cite{Starobinsky:1986fx}:
\begin{align}
  \frac{d\varphi}{dt}=-\frac{V'}{3H}+\frac{H^{3/2}}{2\pi}\eta,\quad \langle\eta\qty(t_1)\eta\qty(t_2)\rangle=\delta\qty(t_1-t_2),
  \label{eq:langevin}
\end{align}
where $\eta$ is the normalized white noise and $\varphi$ is considered a classical stochastic variable.

Due to the effect of classicalization, the quantum correlation of the long wavelength fluctuations becomes classical; therefore, we can treat the probability distribution $P\qty(\varphi,t)$ as the inflaton state.
The quantum state $\hat\rho_\text{long}$ of the inflaton field in the h-region is expressed using the Wigner function
\begin{align}
    W\qty(\varphi,\pi_\varphi)=\int\frac{d\Delta}{2\pi}\bra{\varphi-\frac{\Delta}{2}}\hat\rho_{\mathrm{long}}\ket{\varphi+\frac{\Delta}{2}}\exp(i\Delta\, \pi_\varphi),\label{dfn:Wigner}
\end{align}
where $\pi_\varphi$ denotes the conjugate momentum of $\varphi$.
Because the Wigner function is the quasi-probability distribution, it can be identified with the probability distribution $P\qty(\varphi)$ of the stochastic approach to inflation.
Moreover, $P(\varphi)$ and $W(\varphi)$ (integrated about $\pi_\varphi$) can be derived from the same master equation;  the Wigner function obeys the Lindblad type master equation (see the Appendix for the derivation) and are related as \cite{Nambu:1991vs}
\begin{align}
   P(\varphi)= \int d\pi_\varphi W\qty(\varphi,\pi_\varphi).
\end{align}
The time evolution of the inflaton is described by the Fokker-Planck equation for $P\qty(\varphi,t)$ 
\begin{align}
  \pdv{P\qty(\varphi,t)}{t}&=\hat{L}_\text{FP} P\qty(\varphi,t),\notag\\\quad\hat{L}_\text{FP}&:=\pdv{\varphi}\qty[\frac{V'}{3H}+\frac{H^{3/2}}{8\pi^2}\pdv{\varphi}H^{3/2}].
  \label{eq:FP}
\end{align}
Using the noise average $\expval{\cdot}_\eta$, we can express $P\qty(\varphi,t)$ as 
\begin{align}
P\qty(\varphi,t)=\langle\delta\qty(\varphi-\varphi_\eta(t))\rangle_\eta,
  \label{equally_weighted}
\end{align}
where $\varphi_\eta(t)$ is the solution to Eq.~\eqref{eq:langevin} without the average noise.
In de Sitter spacetime, the number of h-regions increases because of the accelerated expansion of the universe.
However, the probability $P\qty(\varphi,t)$ does not include the global property of the inflationary universe.
Therefore, to consider the global spatial distribution of $\varphi$, we should use the volume-weighted probability distribution defined as \cite{Nakao:1988yi}
\begin{align}
  P_V\qty(\varphi,t)=\dfrac{\displaystyle{\biggl<\delta\qty(\varphi-\varphi_\eta(t))\exp(3\!\!\int^t dt \,H\qty(\varphi_\eta(t)))\biggr>_\eta}}{\displaystyle{\biggl<\exp(3\!\!\int^tdt\,H \qty(\varphi_\eta(t)))\biggr>_\eta}}.
\end{align}
The time evolution equation of this probability distribution $P_V\qty(\varphi,t)$ is described by the Fokker-Planck equation
\begin{equation}
    \pdv{P_V\qty(\varphi,t)}{t}=\hat{L}_\text{FP}P_V\qty(\varphi,t)+3\qty(H-\langle H\rangle_V)P_V\qty(\varphi,t) ,\label{modified_FP_Eq}
\end{equation}
where $\langle X \rangle_V=\int\dd{\varphi}X\qty(\varphi)P_V\qty(\varphi,t)$. 
Here, the last two terms express the non-Markovian nature of stochastic inflation.  The non-Markovian effect is due to the inhomogeneous expansion effect of the universe because the volume of an h-region depends on its initial value and the evolutionary history of the inflaton's trajectory.

\subsection{Equivalence of EE and Shannon Entropy}
For a continuous probability distribution $P(\varphi)$, it is possible to assign Shannon entropy (information entropy). By discretizing $P(\varphi)$, a probability is defined as
\begin{equation}
 P_j:=\mathrm{Pr} (\varphi_j\leq\varphi\leq \varphi_j+\Delta\varphi),
\end{equation}
with normalization $1=\sum_j P_j=\sum_j P(\varphi_j)\Delta\varphi$. Shannon entropy is defined as 
\begin{equation}
S_\text{Sh}:=-\sum_j P_j\ln P_j.
\end{equation}
In the continuous limit $\Delta\varphi\rightarrow 0$,
\begin{equation}
 S_\text{Sh}[P]=-\int d\varphi P(\varphi)\ln P(\varphi),
 \label{eq:wig}
\end{equation}
where we omit the term $-\lim_{\Delta\varphi\rightarrow 0}(\Delta\varphi\ln\Delta\varphi)$ because we are interested only in the entropy differences of the Shannon entropy.
Assuming that the quantum state for stochastic inflation is Gaussian, the state of an h-region can be specified by the following covariance matrix
\begin{equation}
 M_A:=\begin{bmatrix}
 \expval{\{\hat\varphi,\hat\varphi\}} &\expval{\{\hat\varphi,\hat\pi_\varphi\}}\\
 \expval{\{\hat\varphi,\hat\pi_\varphi\}} &\expval{\{\hat\pi_\varphi,\hat\pi_\varphi\}}
 \end{bmatrix},
\end{equation}
where $\hat\varphi$ and $\hat\pi_\varphi$ are local operators assigned to the h-region which are defined by the inflaton field operator by coarse grinding. Introducing a vector $\bm{\xi}=(\varphi,\pi_\varphi)^T$ in the phase space,
the state is expressed using the Wigner function as 
\begin{align}
W(\varphi,\pi_\varphi)&=\frac{1}{\sqrt{\mathrm{det}M_A }\,\pi}\exp\left(-\bm{\xi}^T M_A^{-1}\bm{\xi}\right)  \notag \\
&=\frac{1}{\pi\nu}\exp\left[-\frac{1}{\nu}\tilde{\bm{\xi}}\,\tilde{\bm{\xi}}^T\right],
\label{eq:gauss-state}
\end{align}
where we have adopted the standard form of the covariance matrix with a symplectic matrix S
\begin{equation}
 SM_AS^{-1}=\mathrm{diag}(\nu,\nu),
\end{equation}
where $\nu\ge 1$ is the symplectic eigenvalue of $M_A$ \cite{Adesso:2004}.
Accordingly, canonical variables are transformed as $\tilde{\bm{\xi}}=S^{-1}\bm{\xi}$.
\footnote{The state \eqref{eq:gauss-state} represents the thermal state. The symplectic eigenvalue is related to the inverse temperature $\beta$ as $\nu=1/\tanh(\beta/2)$. $\nu=1$ correspond to the pure state ($\beta=\infty$).} Regarding \eqref{eq:gauss-state} as a probability distribution in the phase space, the Shannon entropy for this state is
\begin{equation}
 S_\text{Sh}=-\int_{-\infty}^{+\infty} d\varphi\, d\pi_\varphi\, W\ln W=\ln(\pi\nu)+1.
 \label{eq:Ssh}
\end{equation}
In contrast, the entanglement entropy (von Neumann entropy) of the Gaussian state \eqref{eq:gauss-state} is \cite{Holevo:2001}
\begin{align}
 S_\text{EE}&=\left(\frac{\nu+1}{2}\right)\ln\left(\frac{\nu+1}{2}\right)-\left(\frac{\nu-1}{2}\right)\ln\left(\frac{\nu-1}{2}\right) \notag \\
 &\approx \ln\nu+1\quad\text{for}\quad \nu \gg 1.
 \label{eq:SEEnu}
\end{align}

The condition of ``classicalization" of the inflaton state implies $\nu\gg 1$; then, the entanglement between the h-region and its complementary region becomes large, whereas the quantum correlation between adjacent h-regions vanishes due to their monogamy property \cite{Nambu2023b}.
The Shannon entropy coincides with the entanglement entropy for $\nu\gg 1$, realized in the stochastic inflation stage.  Therefore, Shannon entropy can be regarded as the entanglement entropy, which quantifies the quantum correlation shared between the h-region and its complementary region.

\section{\label{sec:dw_pte_inf}Inflation with Double-well potential}
Here, we consider an inflationary model with a double-well inflaton potential:
\begin{align}
  V\qty(\varphi)=\frac{\lambda}{4}\qty(\varphi^2-\frac{m^2}{\lambda})^2.
\end{align}
Then, the slow-roll condition is described as \cite{Sasaki:1987gy} 
\begin{align}
    \frac{m^4}{m_p^4} \ll \lambda \ll \frac{m^2}{m_p^2} \ll 1.
    \label{slow-roll condition}
\end{align}
By introducing dimensionless variables $x,u$ defined by
\begin{align}
  \varphi=\frac{m}{\sqrt{\lambda}}x,\quad t=\frac{3H_0}{m^2}u,
\end{align}
with $H_0^2=(\kappa/3)V\qty(0)=\kappa \,m^4/(12\lambda)$,
the Langevin equation \eqref{eq:langevin} becomes
\begin{align}
\begin{split}
  \frac{dx}{du}
  &= x\;\mathrm{sign}\qty(1-x^2)
  + \sqrt{2\alpha}\abs{1-x^2}^{3/2}\eta,\\
  &\langle\eta\qty(u_1)\eta\qty(u_2)\rangle
  = \delta\qty(u_1-u_2),
  \label{dimless_eom}
\end{split}
\end{align}
where a dimensionless parameter is introduced as
\begin{align}
  \alpha=\frac{3\lambda}{8\pi^2}\qty(\frac{H_0^2}{m^2})^2=\frac{\kappa^2}{8\pi^2}\frac{m^4}{48\lambda}.
\end{align}
The dimensionless parameter expresses the strength of the noise from the sub-horizon mode.
The physical volume of a comoving region is then expressed as 
\begin{align}
  \mathscr{V}
  &=\mathscr{V}_0 \exp(3\int^t\dd{t}H\qty(\varphi\qty(t)))\notag\\
  &= \mathscr{V}_0\exp(\gamma\int^u\dd{u}\abs{1-x^2(u)}),
\end{align}
where $\mathscr{V}_0$ denotes the comoving volume and the second dimensionless parameter is introduced by
\begin{equation}
\gamma=\frac{3H_0^2}{m^2}
=\frac{\kappa\, m^2}{4\lambda}.
\end{equation}

We choose the initial value of the inflaton field as $x_0\ll 1$.
The slow-roll condition in \eqref{slow-roll condition} is re-expressed as
\begin{align}
  \alpha\ll 1,\quad \gamma\gg 1.
  \label{alpha_gamma}
\end{align}
The value of the dimensionless inflaton field $x_\text{end}$ at the end of the slow-roll phase is derived from \eqref{eq:inflaton_end_time} using $\gamma\gg 1$:
\begin{align}
    x_\text{end}\approx 1-\frac{1}{2\sqrt{\gamma}}.
\end{align}
The total e-folding number is
\begin{equation}
\mathcal{N}_e=\int_{t_0}^{t_\text{end}}d\varphi\frac{H}{\dot\varphi}\approx\gamma\ln\left(\frac{x_\text{end}}{x_0}\right)
\approx\gamma\ln\left(\frac{1}{x_0}\right).
\end{equation}
Then, the curvature fluctuation at $t_0$ is expressed as 
\begin{align}
    \frac{\langle\mathcal{R}^2\rangle_{t_0}}{2}
    &=\left.\frac{\kappa^3}{24\pi^2}\frac{V^3}{(V')^2}\right|_{t_0}\notag\\
    &=\alpha\,\gamma\frac{(x_{0}^2-1)^4}{x^2_{0}}\approx \frac{\alpha \gamma}{x_0^2}.
\end{align}
 The Gibbons-Hawking entropy \eqref{eq:GH} is
\begin{equation}
 S_\text{GH}(t_\text{end})=\frac{1}{4\alpha(1-x_\text{end}^2)^2}\approx\frac{\gamma}{4\alpha}.
\end{equation}
We examine the inequality \eqref{eq:entropy_bound} introduced by \cite{Arkani-Hamed:2007ryv}. The value of the curvature fluctuation is
\begin{equation}
\expval{\mathcal{R}^2}_{t_0}=\frac{2\alpha\gamma}{\expval{x_0^2}}=\gamma,
\end{equation}
where we have assumed that the variance of the initial value $x_0$ is the same as fluctuation of the quantum noise $2\alpha$. Thus the inequality becomes
\begin{equation}
\frac{2\mathcal{N}_e}{\gamma}<S_\text{GH}(t_\text{end}),
\label{eq:ineq0}
\end{equation}
where the left-hand side is evaluated to be  $2S_\text{Sh}(t_\text{end};t_0)$ as explicated by \eqref{Late_time_limit_Non_volume_weight}, and incorporates the stricter inequality compared to \eqref{entropy_bound_precise}.

From~\eqref{entropy_bound_precise}, our proposal of entropy bound for de Sitter spacetime is
\begin{equation}
 S_\text{Sh}(t_\text{end};t_0)\leq S_\text{GH}(t_\text{end}).
\end{equation}
This inequality includes inequality \eqref{eq:ineq0}. 
To determine the left-hand side of this inequality, we should obtain the probability distribution function of the inflaton field.

\subsection{No volume-weighted case (local picture)}
As we assumed Gaussian form for the Wigner function, we also assume the solution of the Fokker-Planck Equation \eqref{eq:FP} is Gaussian form
\begin{align}
    P(x,u)=\frac{1}{\sqrt{2\pi \sigma_0(u)}}\mathrm{exp}
    \left[
        -\frac{ (x-x_{\mathrm{c}}(u))^2}{2\sigma_0(u)}    
    \right], \label{eq:Pxu}
\end{align}
where the average $\langle x\rangle =x_{\mathrm{c}}(u)=x_0e^{u}$ is the classical solution of the dimensionless inflaton's equation of motion \eqref{dimless_eom}.
From this assumption, 
we have to determine the initial dispersion $\sigma_0(0)$.
We can estimate the initial dispersion of the state assuming the equilibrium of the inflaton with thermal Hawking radiation with temperature $T_H=H_0/(2\pi)$. In this case, the initial dispersion of the inflaton is
$\langle \varphi^2(0)\rangle \approx H^2_0$. In terms of the dimensionless variables $x$, the dispersion satisfies $\langle x^2\rangle -\langle x\rangle^2=\sigma_0(0)=\alpha/\gamma$. Consequently, the dispersion is given by
\begin{equation}
\sigma_0(u)=\left(\alpha+\frac{\alpha}{\gamma}\right)e^{2u}-\alpha.
\label{eq:sig0}
\end{equation}

\begin{figure}[b]
\centering
\includegraphics[width=0.9\linewidth]{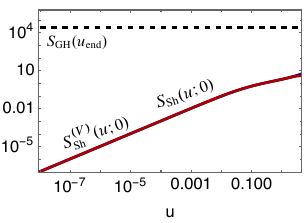}
\caption{
Entropy behavior for the non-eternal inflation case ($\alpha=1.0\times 10^{-4}, \gamma=10, x_0=\sqrt{2\,\alpha}, u_\text{end}=4.1$). The total e-folding number is $41$. The entropy $S_{\text{Sh}}$ and $S^{(V)}_\text{Sh}$ have the same behavior and do not violate the entropy bound.  }
\label{fig:ssh_sgh_a=1e-5_g=40_x_0=3000lam}
\end{figure}

The Shannon entropy for the inflaton field with the probability distribution \eqref{eq:Pxu} is 
\begin{align}
  S_{\mathrm{Sh}}(u;u_0)&=S_\text{Sh}[P(u)]-S_\text{Sh}[P(0)]\notag\\
  &=\frac{1}{2}\ln(\frac{\sigma_0(u)}{\sigma_0(0)}).
\end{align}
\begin{figure}[t]
\centering
\includegraphics[width=0.9\linewidth]{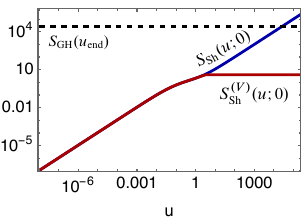}
\caption{
Entropy behavior for the eternal inflation case $(\alpha=1.0\times 10^{-4}$, $\gamma=10$, $x_0=e^{-2\gamma/\alpha}, u_\text{end}=2.0\times 10^5)$. The total e-foling number is $2.0\times 10^6$. $S_{\text{Sh}}$ (blue line) exceeds the entropy bound (dotted line) whereas $S_{\text{Sh}}^{(V)}$ (red line) remains within the  bound.}
\label{fig:ssh_sgh_sv_a=1e-5_g=40}
\end{figure}
The late-time $u\gg 1$ behavior of the Shannon entropy is 
\begin{align}
  S_{\mathrm{Sh}}(u;u_0)\sim u =\ln\left(\frac{x_c(u)}{x_0}\right)=
  \frac{1}{\gamma}(N-N_0).
  \label{Late_time_limit_Non_volume_weight}
\end{align}
As the dimensionless variable $u$ is proportional to the e-folding number $N-N_0$, we confirmed that Shannon entropy and entanglement entropy of a single h-region are proportional to the e-folding number of inflation, which establishes the operational meaning of the entropy bound~\cite{Arkani-Hamed:2007ryv} in the stochastic approach to inflation in terms of entanglement entropy (Shannon entropy) and GH entropy.

From the Langevin equation~\eqref{dimless_eom}, the initial value $x_0$ controls whether inflation enters the eternal phase or not.
When eternal inflation occurs, the random force dominates the potential force at the initial value. 
Thus, the condition of eternal inflation is 
\begin{align}
  x_0 < \sqrt{2\,\alpha}.
  \label{eq:Eti_cond}
\end{align}
In \cite{Arkani-Hamed:2007ryv}, because the e-folding number violates the entropy bound during the eternal inflation, they concluded that eternal inflation is in the swampland.
Therefore, we set $x_0 < \sqrt{2\,\alpha}$ to confirm that the entropy bound is violated in the case of eternal inflation. 
To compare the eternal inflation case, we also consider the non-eternal inflation case: $\sqrt{2\,\alpha}\leq x_0$.
The time evolution of the entropy is shown in Fig.~\ref{fig:ssh_sgh_a=1e-5_g=40_x_0=3000lam} (non-eternal case) and Fig.~\ref{fig:ssh_sgh_sv_a=1e-5_g=40} (eternal case).
From these, we observe that Shannon entropy $S_{\text{Sh}}\qty(P)$ does not violate the entropy bound \eqref{entropy_bound_precise} for the non-eternal inflation case but violates the entropy bound for the eternal inflation case.
These behaviors correspond to the result presented in \cite{Arkani-Hamed:2007ryv}.


We recognize this violation of the entropy bound as an information problem. 
If the entropy bound is violated, the observer can extract the information of the h-region without a limit, and any excess information will exist in the super-horizon modes of the inflaton field.
This also indicates a violation of the system's unitarity.
Therefore, from a quantum information perspective, this problem should be resolved similarly to the black hole information loss paradox.
For this, we require guiding behavior of entropy, such as the Page curve in the black hole information loss paradox context.
We focus on the expansion effect of the universe as a typical nature in the de Sitter spacetime to gain guiding behavior. 
The expansion effect is related to the volume of the universe; therefore, we should consider the volume-weighted probability distribution to express a typical state in the de Sitter spacetime. 
We discuss the behavior of entropy using the volume-weighted probability distribution in the next subsection.

\subsection{Volume-weighted case (global picture)}
The solution of the Fokker-Planck equation for volume-weighted probability \eqref{modified_FP_Eq} is derived approximately as \cite{Nakao:1988yi}
\begin{align}
  &P_V\qty(x,u)=\sqrt{\frac{1+g(u)\,\sigma_V(u)}{2\pi\sigma_V(u)}}\notag \\
  &\times\exp[-\frac{1+g(u)\,\sigma_V(u)}{2\sigma_V(u)}\qty(x-\frac{x_c(u)}{1+g(u)\,\sigma_V(u)})^2].
  \label{Sol_Vol_FP.Eq}
\end{align}
Here, $\sigma_V\qty(u)$ and $g\qty(u)$ are obtained on the basis of the WKB approximation $\qty(\alpha\ll 1)$ as
\begin{align}
  &\sigma_V\qty(u)=\sigma_0(u)\notag\\
  &\quad +3\gamma\alpha^2\qty[\frac{D}\alpha\qty(1-e^{2u}+2ue^{2u})+1+2u-e^{2u}],\\
  &g\qty(u)=3\gamma\qty[\frac{D (e^{2u}-1)-2\alpha u}{\sigma_0(u)}]+\mathcal{O}\qty(\alpha^2),
\end{align}
where $D=\alpha+\alpha/\gamma$ and we impose assumptions $g(0)=0$ and $\sigma_V(0)=\sigma_0(0)$ to set the initial distribution equal to the non-weighted probability distribution \eqref{eq:Pxu}.
Therefore, the Shannon entropy of $P_V$ is 
\begin{align}
&S_{\mathrm{Sh}}^{(V)}(u;0)
=S_{\mathrm{Sh}}[P_V(u)]-S_{\mathrm{Sh}}[P_V(0)]\notag\\
&=\frac{1}{2}\ln\qty(\frac{1+g(u_0)\,\sigma_V(u_0)}{\sigma_V(u_0)}\frac{\sigma_V(u)}{1+g(u)\,\sigma_V(u)}).
\end{align}
Subsequently, the entropy approaches a constant value:
\begin{align}
  S_{\mathrm{Sh}}^{(V)}(u;0)\sim \frac{1}{2}\ln\qty(\frac{1}{3\gamma D})=\text{const.} \quad \qty(u\gg1)
  \label{Late_time_limit_volume_weight}
\end{align}


From the Fokker-Planck equation~\eqref{modified_FP_Eq} and its solution \eqref{Sol_Vol_FP.Eq}, the trajectory of the peak of the probability distribution differs from the classical trajectory of the inflaton field.
Unlike $P\qty(\varphi,t)$, the standard deviation of $P_V$ remains finite. 
This difference originates from the volume term in the Fokker-Planck equation~\eqref{modified_FP_Eq}.
Because of this difference, two distinct phases, namely eternal and non-eternal inflation phases, occurred in the setup.
The difference in these phases comes from the duration of inflation.

If we consider the volume-weighting for the probability distribution, the probability of the inflaton state increases owing to the volume effect.
Then, the peak of the probability distribution climbs up the potential slope because the state with the higher potential value has a larger volume.
We can neglect the contributions of states with 
smaller volumes, preventing the probability distribution from reaching the bottom of the potential.
Thus, the Shannon entropy approaches a constant value in the late time if the volume-weighting effect is dominant.
From this, the condition under which eternal inflation occurs is described by \cite{Nakao:1988yi}
\begin{align}
    x_0^2 \ll 3\alpha\gamma.
    \label{eq:eternal}
\end{align}
Compared to the condition \eqref{eq:Eti_cond}, the condition \eqref{eq:eternal} is less restrictive, and the dominance of quantum fluctuations is sufficient to sustain eternal inflation in the present potential.

The evolution of the entropy $S_\text{Sh}^{(V)}$ is shown in Fig. \ref{fig:ssh_sgh_a=1e-5_g=40_x_0=3000lam} for the non-eternal phase and in Fig. \ref{fig:ssh_sgh_sv_a=1e-5_g=40} for the eternal phase. We observe that the Shannon entropy with volume-weighting in the eternal inflation case does not violate the entropy bound \eqref{entropy_bound_precise}.
In the volume-weighting case, we account for the contribution of h-regions with smaller Hubble radii. This does not imply considering subhorizon-scale fluctuations; rather, it represents a form of “fine-graining,” as a larger Hubble parameter corresponds to a smaller spatial size. Conversely, in the probability distribution without volume-weighting, the contribution from h-regions with smaller Hubble radii is ignored. This neglect corresponds to “coarse-graining” the effective degrees of freedom.

At late times, the entropy of the volume-weighted probability distribution becomes stationary, indicating that an equilibrium state is achieved in terms of $P_V(\varphi)$. On the other hand, as time progresses, superhorizon modes become entangled with their complementary degrees of freedom. This increasing entanglement reflects the effect of fine-graining, which reduces the entropy.

\section{\label{sec:sum_conc}Conclusion}

We confirmed that an information problem exists in the de Sitter spacetime as in the black hole spacetime. 
Considering the typicality of the quantum states, we obtain a more meaningful behavior of the entropy that maintains the entropy bound.
For volume-weighted probability, the Shannon entropy becomes constant in late time, and the entanglement of a single h-region no longer increases. 
Thus, the inflaton field attained an equilibrium state in the late time.
However, the details of the information dynamics in our setup should be discussed in greater detail.

In our calculation, the stochastic process becomes a non-Markovian process owing to the volume-weighting.
The observer can observe the h-regions at the end time of inflation and those under the inflationary period.
This property is included as the nonlocal effect due to volume weighting.
This probability treatment is similar to that of the island formula \cite{Almheiri:2019hni, Penington:2019kki, Penington:2019npb, Almheiri:2020cfm} in black hole evaporation.
Therefore, we conclude that the global effect reflecting the backreaction of the quantum fluctuations of the inflaton field on the geometry is included in the stochastic approach to inflation.

From a quantitative perspective, the influence of volume-weighting is pronounced when the probability distribution coincides with the apex of the potential. Consequently, our framework can be extended to encompass other potential configurations, such as those present in multi-field inflation models. Within the realm of string theory, the de Sitter swampland conjecture emerges from the distance conjecture~\cite{Ooguri:2018wrx}. This conjecture suggests that slow-roll inflation contradicts string theory and has connections to the entropy bound~\cite{obied2018sitter}. This conjecture is also evaluated in the context of eternal inflation~\cite{Matsui:2018bsy}. To reconcile an inflationary model with this conjecture, a stochastic approach to eternal inflation is explored \cite{Fanaras:2023acz}, excluding volume-weighting. Future research should focus on applying our methods to this conjecture to probe the resolution of the discrepancies between the inflationary paradigm and string theory.

\begin{acknowledgments}
YN was supported by JSPS KAKENHI (Grant No.~JP23K25871) and MEXT KAKENHI Grant-in-Aid for Transformative Research Areas A ``Extreme Universe"(Grant No.~24H00956).
\end{acknowledgments}

\appendix
\section{Derivation of Lindblad master equation in stochastic approach}
Assuming that the potential has only a mass term
\begin{align}
    V_2\qty(\varphi)=\frac{1}{2}m^2\varphi^2,
\end{align}
we derive the Lindblad master equation from the following Fokker-Planck equation for the Wigner function $W(Q,P,t)$ in the phase space~\cite{Nambu:1991vs}:
\begin{align}
\pdv{W}{t}=&\qty[-\frac{P}{a^3}\pdv{Q}+m^2a^3Q\pdv{P}]W\notag\\
&+\frac{H^3}{8\pi^2}\qty[\pdv{Q}+\frac{m^2a^3}{3H}\pdv{P}]^2W.\label{Eq_Wigner}
\end{align}
First, we rewrite the first term of the density operator. From the definition of the Wigner function \eqref{dfn:Wigner}, 
\begin{align}
    P\frac{\partial W}{\partial Q}
    =\int\dd{\Delta}P\, e^{2iP\Delta}\pdv{Q}\mel{Q-\Delta}{\,\hat{\rho}\,}{Q+\Delta},
\end{align}
where $P$ is the c-number. Therefore, we obtain 
\begin{align}
  &P\frac{\partial W}{\partial Q}
  =\frac{1}{2}\int\dd{\Delta}\pdv{\Delta}\biggl[e^{2iP\Delta}\mel{Q-\Delta}{\qty[\hat P,\hat\rho]}{Q+\Delta}\biggr] \notag\\
  &\quad - \frac{1}{2}\int\dd{\Delta}e^{2iP\Delta}\pdv{\Delta}\mel{Q-\Delta}{\qty[\hat P,\hat\rho]}{Q+\Delta},
\end{align}
where $\hat P:=-i\partial_Q$.
Then, without loss of generality, we assume the Gaussian state and the first term in the \eqref{Eq_Wigner} is rewritten as
\begin{align}
  &-\frac{P}{a^3}\frac{\partial W}{\partial Q}\notag \\
  &=-i\int d\Delta \,e^{2iP\Delta}\mel{Q-\Delta}{\qty[\frac{\hat{P}^2}{2a^3},\hat\rho]}{Q+\Delta}.
\end{align}
Then, doing the same calculation for the second term in the \eqref{Eq_Wigner} , we obtain
\begin{align}
  &\qty[-\frac{P}{a^3}\pdv{Q}+m^2a^3Q\pdv{P}]W \notag \\
  &=
  -i\int d\Delta \,e^{2iP\Delta}\langle Q-\Delta|\left[\hat{\mathcal{H}},\hat{\rho}\right]|Q+\Delta\rangle,
\end{align}
where
\begin{equation}
\hat{\mathcal{H}}=\frac{\hat{P}^2}{2a^3}+\frac{a^3m^2}{2}\hat Q^2.
\end{equation}
The remaining term in \eqref{Eq_Wigner} can be written as
\begin{align}
    &\qty(\pdv{Q}+\frac{m^2a^3}{3H}\pdv{P})^2W
   \notag  \\&=-\int\dd{\Delta}e^{2iP\Delta}\qty[\hat{L}\hat{\rho}\hat{L}-\frac{1}{2}\qty{\hat{L}^\dagger \hat{L},\hat{\rho}}],
\end{align}
where the bracket $\qty{~,~}$ expresses the anti-commutator, and the operator $\hat{L}$ is defined by
\begin{align}
      \hat{L}:=\sqrt{\frac{H^3}{4\pi^2}}
  \qty(\hat{P}-\frac{m^2a^3}{3H}\hat{Q})=\hat{L}^\dagger.
\end{align}

Finally, we obtain the Lindblad equation for stochastic inflation:
\begin{align}
\frac{\partial\hat\rho}{\partial t}=-i\qty[\hat{\mathcal{H}},\hat\rho]+\hat{L}\hat{\rho}\hat{L}^\dag-\frac{1}{2}\qty{\hat{L}^{\dagger}\hat{L},\hat{\rho}}.
\end{align}
For the general form of the potential $V(\hat{Q})$, we make the following replacement in the Lindblad operator $\hat{L}$: $
    a^3m^2\hat{Q}/(3H)\to a^3 V'(\hat{Q})/(3H)$,
and in the Hamiltonian $\hat{\mathcal{H}}$:
$m^2\hat Q^2/2\to V(\hat Q)$.

\bibliography{Draft.bib}

\end{document}